\documentclass[aps,prl,groupedaddress,twocolumn,10pt]{revtex4-1}
\usepackage{graphicx}
\usepackage{epstopdf}
\usepackage[english]{babel}
\usepackage[T1]{fontenc}
\usepackage{verbatim}
\usepackage{float}

\usepackage{makecell}
\usepackage{color}
\usepackage{multirow}
\usepackage{booktabs}
\usepackage{amsmath}
\usepackage{amssymb}
\usepackage{amsthm}
\usepackage{bbm}
\usepackage{hyperref}
\usepackage[T1]{fontenc}
\usepackage[usenames,dvipsnames]{xcolor}
\hypersetup{colorlinks=true, linkcolor=blue, urlcolor=blue!50!black, citecolor=red!50!black}

\usepackage[normalem]{ulem}

\renewcommand{\imath}[0]{\mathsf{i}}

\usepackage[nodayofweek,level]{datetime}

\begin{document}

\title{Vesiculation mechanisms mediated by anisotropic proteins}
\author{Ke Xiao}
\email{xiaoke@ucas.ac.cn}
\affiliation{Wenzhou Institute, University of Chinese Academy of Sciences, Wenzhou 325016, People's Republic of China}
\affiliation{Department of Physics, College of Physical Science and Technology, Xiamen University, Xiamen 361005, People's Republic of China}
\author{Chen-Xu Wu}
\email{cxwu@xmu.edu.cn}
\affiliation{Department of Physics and Fujian Provincial Key Laboratory for Soft Functional Materials Research, College of Physical Science and Technology, Xiamen University, Xiamen 361005, People's Republic of China}
\author{Rui Ma}
\email{ruima@xmu.edu.cn}
\affiliation{Department of Physics and Fujian Provincial Key Laboratory for Soft Functional Materials Research, College of Physical Science and Technology, Xiamen University, Xiamen 361005, People's Republic of China}

\begin{abstract}
Endocytosis is an essential biological process for the trafficking of macromolecules (cargo) and membrane proteins in cells. In yeast cells, this involves the invagination of a tubular structure on the membrane and the formation of endocytic vesicles. Bin/Amphiphysin/Rvs (BAR) proteins holding a crescent-shape are generally assumed to be the active player to squeeze the tubular structure and pinch off the vesicle by forming a scaffold on the side of the tubular membrane. Here we use the extended Helfrich model to theoretically investigate how BAR proteins help drive the formation of vesicles via generating anisotropic curvatures. Our results show that, within the classical Helfrich model, increasing the spontaneous curvature at the side of a tubular membrane is unable to reduce the tube radius to a critical size to induce membrane fission. However, membranes coated with proteins that generate anisotropic curvatures are prone to experience an hourglass-shaped necking or a tube-shaped necking process, an important step leading to membrane fission and vesicle formation. In addition, our study shows that depending on the type of anisotropic curvatures generated by a protein, the force to maintain the protein coated membrane at a tubular shape exhibits qualitatively different relationship with the spontaneous curvature. This result provides an experimental guidance to determine the type of anisotropic curvatures of a protein.
\end{abstract}
\date{\today}

\maketitle

Endocytosis is involved in many cellular processes, including nutrient uptake, regulated recycling of plasma membrane components, and neural signaling~\cite{W.Kukulski2012}.
This process is achieved through a formation of transient, highly curved membrane configurations such as tubules or vesicles, which have the targeted molecules wrapped  inside~\cite{J.S.Bonifacino2004}.
During endocytosis  in yeast cells, a small patch of the plasma membrane is first deformed into a shallow invagination, which is subsequently elongated into a deep one, followed by a constriction of its neck until a cargo-carrying vesicle is formed and pinched off~\cite{T.Kishimoto2011}.
These membrane-shaping events are generally mediated by a plethora types of proteins bound to the membrane~\cite{M.J.Taylor2011,A.Mahapatra2021,D.Perrais2005,H.T.McMahon2011}.
The presence of different types of proteins on the membrane gives rise to changes in mechanical properties of the membrane, such as bending rigidity~\cite{A.F.Loftus2012} and membrane curvature~\cite{I.Tsafrir2001}.
Clathrin proteins assemble into a lattice with a mixture of pentagons and hexagons which scaffold the flat membrane into a spherical shape~\cite{O.Avinoam2015,G.Kumar2016,D.Bucher2018,Z.M.Chen2020}. The GTPase dynamin proteins form a helical band at the neck of the endocytic pit. It is generally thought that the constriction of the band upon GTP hydrolysis drives vesicle scission.
Another active participant to facilitate vesicle scission is the Bin/amphiphysin/Rvs (BAR) domain proteins that are found to be bound at the side of the endocytic pit and assemble into a cylindrical scaffold. The crescent-shaped BAR proteins are expected to bend the membrane into different curvatures in parallel with and in perpendicular with their orientations~\cite{T.Kishimoto2011,C.T.Lee2021,M.Simunovic2015,M.Simunovic2016}. Such a mechanical feature, enhanced by the enrichment of BAR proteins on the membrane, is able to induce tubulation~\cite{A.Frost2008,B.Habermann2004,C.Mim2012}.  The role of BAR proteins as a facilitator for vesicle scission has been challenged by Walani {\it et al.} who proposed that the BAR proteins actually help stabilize the tubular endocytic pit and it is the depolymerization of BAR proteins that leads to the scission of the tubular pit through a snap-through transition induced by high membrane tension~\cite{N.Walani2015PNAS}.

The physical mechanisms behind vesiculation during endocytosis are actively studied both in the context of cell biology and biophysics. Rapid developments in imaging technologies such as electron microscopy and fluorescence microscopy have demonstrated the shapes of endocytic pit at different stages of endocytosis\cite{T.F.Roth1964,R.G.W.Anderson1977,J.Heuser1980}.
Experiments have confirmed that vesicle formation can arise from the conical shape of lipid molecules~\cite{J.C.Stachowiak2013,M.Pinot2014}, membrane tension induced by external forces~\cite{S.Zheng2014,A.Diz-Munoz2013,P.J.Wen2016}, as well as spontaneous curvature generated by membrane-bound proteins~\cite{M.G.J.Ford2002,J.C.Stachowiak2012,D.J.Busch2015,R.Lipowsky2013}.
Theoretical modeling was exploited to interpret experimental observations in a mechanistic context, offering valuable insights into the underlying mechanical principles of membrane budding phenomena~\cite{P.Sens2004,J.Liu2006,N.Walani2015PNAS,T.Zhang2015,J.E.Hassinger2017,M.Rui2021}. Most of these works focus on the explanation of how a flat membrane is deformed into either a tubular pit via external forces or a spherical vesicle via clathrin assembly. The very last step of vesicle scission has been studied only in a limited number of works.  By constructing a quantitative model, Liu {\it et al.}~\cite{J.Liu2006} suggested that the line tension at the interface between different lipid domains on the invaginating membrane is sufficient for a successful vesicle scission during endocytosis. However, experimental evidence for lipid phase separation on the endocytic pit is still lacking.

In the classical Helfrich model of membrane, the effect of curvature generation by proteins on the membrane is embedded in a parameter so called spontaneous curvature. In order to reduce the energetic cost for bending, membrane tends to deform in such a way that the mean curvature of the membrane equals to the spontaneous curvature. However,
this description of curvature generation cannot capture the effect of anisotropic proteins, such as BAR proteins, which tend to bend the membrane independently into different curvatures in different tangential directions. Whether the anisotropic curvature generated by BAR-proteins at the side of a tubular membrane is able to induce vesicle scission remains unclear.

In this paper, using the extended-Helfrich model developed to account for the anisotropic spontaneous curvatures generated by anisotropic proteins~\cite{A.Iglic2005,D.Kabaso2012, N.Bobrovska2013,N.Walani2014PRE}, we investigate how a tubular membrane is deformed by anisotropic proteins bound to the side of the membrane. It is found that within the classical Helfrich model of membrane, increasing the spontaneous curvature cannot lead to membrane fission. Anisotropic spontaneous curvatures are necessary to narrow the membrane into a tubular neck or an hourglass neck. We also suggest an experimental method to distinguish the type of anisotropic spontaneous curvatures generated by a protein by comparing the force to maintain the membrane at a tubular shape in the presence and absence of the protein coat.

\begin{figure}[htp]
\centering
\includegraphics[width=\linewidth,keepaspectratio]{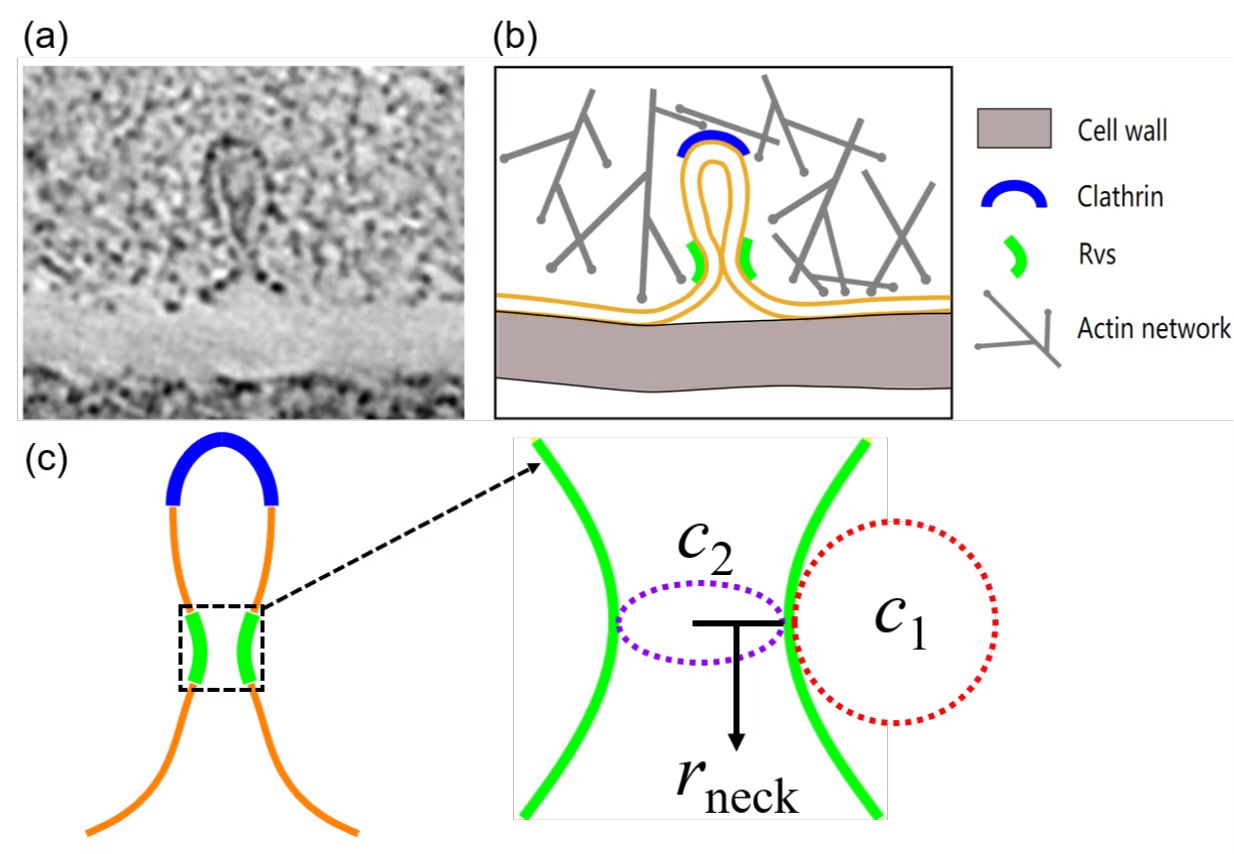}
\caption{(a) Electron micrograph of an endocytic invagination in yeast cells when BAR proteins are present. The graph is adapted from Ref.~\cite{W.Kukulski2012}. (b) Schematic illustration of the proteins involved in endocytosis. The BAR proteins Rvs (red) are bound to the side of the membrane and expected to generate anisotropic curvatures to facilitate membrane fission. (c) A schematic picture of the membrane morphology depicts $r_{\rm neck}$, $c_1$, and $c_2$.
}
\label{fig:electron}
\end{figure}

We consider the deformation of a tubular invagination in the late stage of endocytosis in yeast cells when BAR proteins are present at the side of the tube, as shown in Fig.~\ref{fig:electron}(a) and (b).
The invagination has been pulled into a tubular shape by actin polymerization forces $f$ against the high turgor pressure $p$ inside of the cell. The membrane tension $\sigma$ is assumed to be small due to the presence of eisosomes which serve as a membrane reservoir.
The curvature generated by the clathrin coat at the tip of the invagination is described by the Helfrich model. The anisotropic curvature generated by the BAR proteins at the side of the invagination is described by the extended-Helfrich model with a bending energy density per unit area given by~\cite{N.Walani2014PRE}
\begin{equation}
\label{eq:fb}
f_b = \frac{\kappa}{2}(c_1-c_0^{1})^2+\frac{\kappa}{2}(c_2-c_0^{2})^2 \notag\\
+\kappa_{12}(c_1-c_0^{1})(c_2-c_0^{2}),
\end{equation}
where $\kappa$ denotes the bending rigidity, $c_{1}$ and $c_{2}$ represent the two principal curvatures of the membrane (see Fig.~\ref{fig:electron}(c)), and $c_0^{1}$ and $c_0^{2}$ denote the preferred curvature imposed by the BAR proteins in the longitudinal direction and in the circumferential direction, respectively.  The coupling constant $\kappa_{12}$ determines the type of curvature generated by proteins, which in general deviates from $\kappa$, corresponding to an anisotropic
curvature model. When $\kappa_{12}=\kappa$, it reduces to the classical Helfrich model.

In order to derive shape equations which govern the morphology of the membrane surface at the endocytosis site, Euler-Lagrange variational methods were performed with respect to the total free energy of the membrane. As a result, the shape equations can be computed via minimizing the total free energy functional under the constraints. The details of the model and the equations are presented in the Supplementary Information. We numerically solve these shape equations in Matlab using the 'bvp4c' solver.

As the effect of the anisotropic proteins bound to the side of the endocytic invagination is described by the spontaneous curvature $c_0^2$ in the circumferential direction, we deliberately vary $c_0^2$ so as to study how the membrane morphology changes accordingly.  The membrane morphology is characterized by the mean neck radius $\langle r_{\rm neck} \rangle $ and the mean of two principal curvatures $\langle c_1 \rangle $ and $\langle c_2 \rangle $ over the anisotropic protein coated area. Vesiculation is indicated by reducing the narrowest neck radius below a critical value of $5\mathrm{nm}$.

\begin{figure}[htp]
\centering
\includegraphics[width=\linewidth,keepaspectratio]{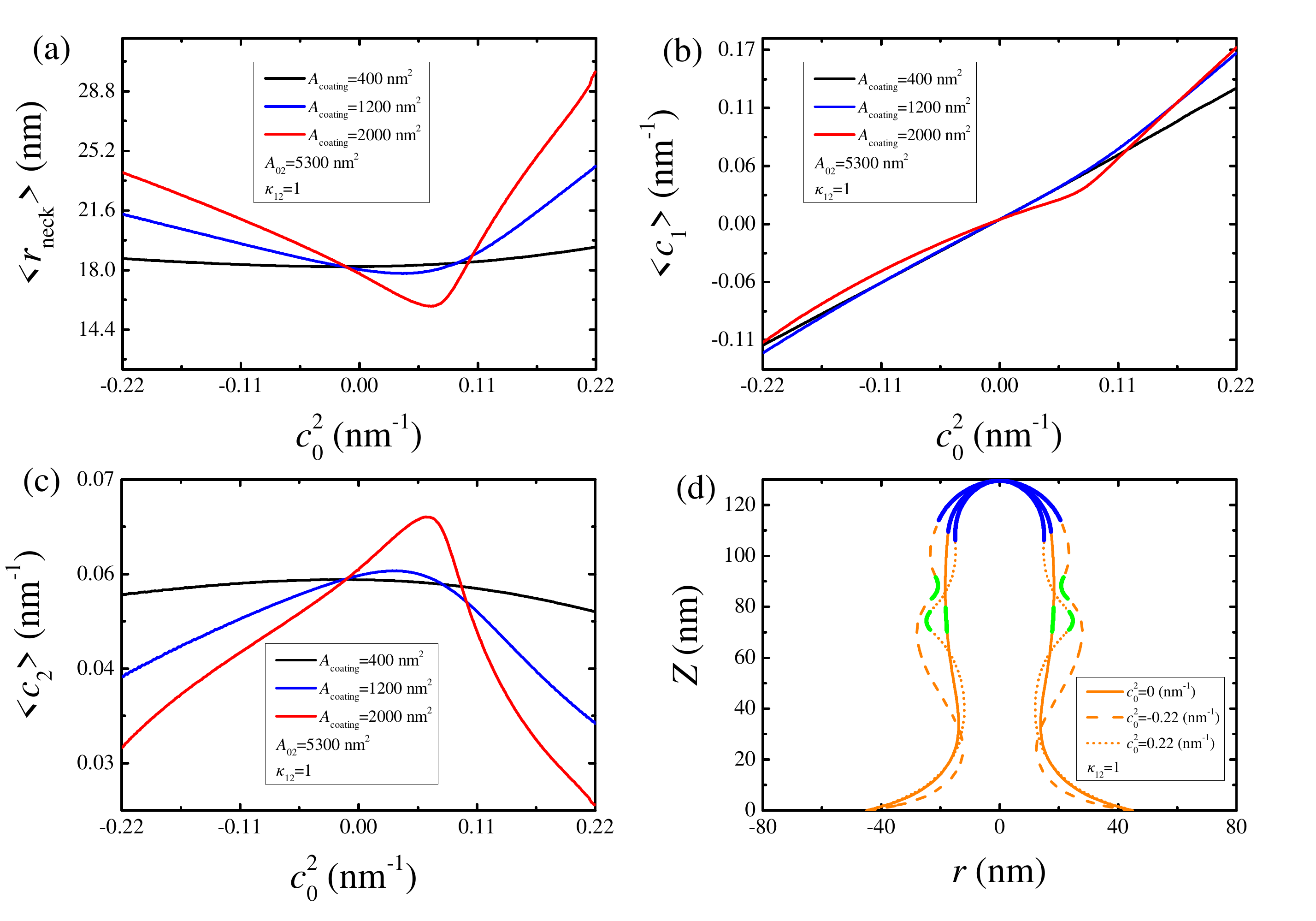}
\caption{The mean value of the neck radius $\langle r_{\rm neck} \rangle $ in (a) and the two principal curvatures $\langle c_1 \rangle $ in (b) and $\langle c_2 \rangle $ in (c), as a function of the spontaneous curvature $c_0^2$ for the Helfrich model $\kappa_{12} = \kappa$. The average is taken over the protein-coated area at the side of the membrane. The three curves in (a-c) correspond to different coating areas. (d) Profile views of membrane morphologies for positive (dotted line), negative (dashed line), and zero (solid line) spontaneous curvature $c_0^2$.
}
\label{RneckVsC02_0}
\end{figure}

For the case of the classical Helfrich model ($\kappa_{12}=\kappa$), varying $c_0^2$ is equivalent to tuning the spontaneous curvature of the model. It is shown that the average neck radius $\langle r_{\rm neck}\rangle$ is a non-monotonic function of $c_0^2$ with a minimum of about $14~\mathrm{nm}$, far from the critical value for vesiculation to occur (5$~\mathrm{nm}$).  The average longitudinal curvature $\langle c_1 \rangle$ increases with $c_0^2$ and changes its sign when $c_0^2$ crosses $0$ (Fig.~\ref{RneckVsC02_0}(b)), while the average circumferential curvature $\langle c_2 \rangle$ reaches a peak value at an intermediate value of $c_0^2$.
Membrane shapes of positive, negative and zero spontaneous curvatures $c_0^2$ are depicted in Fig.~\ref{RneckVsC02_0}(d), where the protein-coated area of positive/negative spontaneous curvatures shows a wider radius than that of zero spontaneous curvature. These results suggest that increasing the spontaneous curvature in the Helfrich model is unable to produce vesiculation.

\begin{figure}[htp]
\centering
\includegraphics[width=\linewidth,keepaspectratio]{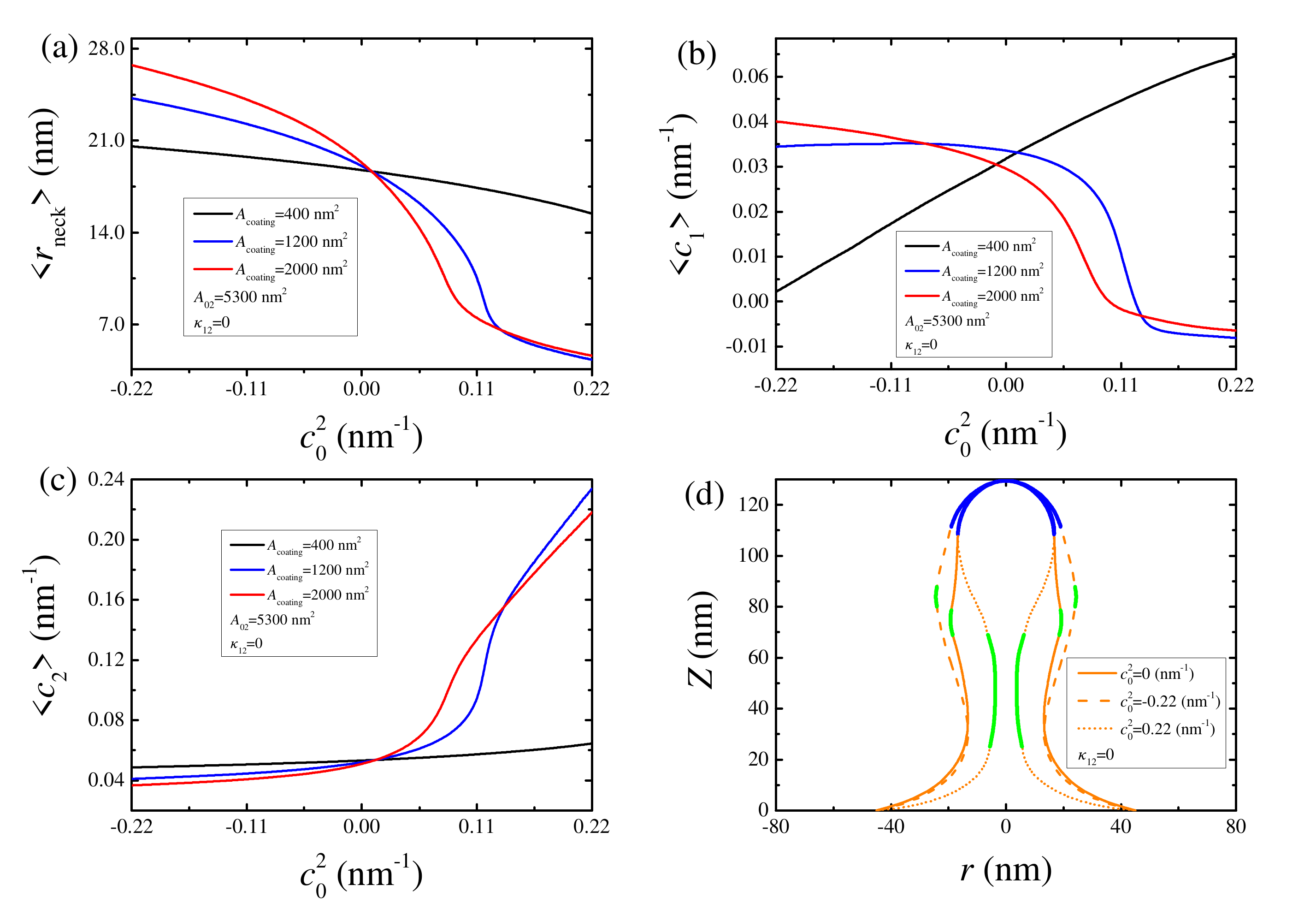}
\caption{The mean value of the neck radius $\langle r_{\rm neck} \rangle $ in (a) and the two principal curvatures $\langle c_1 \rangle $ in (b) and $\langle c_2 \rangle $ in (c), as a function of the spontaneous curvature $c_0^2$ for $\kappa_{12} = 0$. The average is taken over the protein-coated area at the side of the membrane. The three curves in (a-c) correspond to different coating areas. (d) Profile views of membrane morphologies for positive (dotted line), negative (dashed line), and zero (solid line) spontaneous curvature $c_0^2$.
}
\label{RneckVsC02_-1}
\end{figure}

We next investigate whether the extended-Helfrich model is able to produce vesiculation. As a first attempt, we consider $\kappa_{12}=0$.
If the area of the anisotropic protein-coated membrane is small, the average neck radius $\langle r_{\rm neck} \rangle$ and the average circumferential curvature $\langle c_2 \rangle$ are almost independent of $c_0^2$ (see the black curves in Fig.~\ref{RneckVsC02_-1}(a) and (c)), while the average longitudinal curvature $\langle c_1 \rangle $ increases with $c_0^2$ in a considerable way (see the black curve in Fig.~\ref{RneckVsC02_-1}(b)).
In contrast, for larger coating area, $\langle r_{\rm neck} \rangle$ is narrowed down with the increase of $c_0^2$ and the longitudinal curvature $\langle c_1 \rangle $ drops to a value that is close to zero (see blue and red curves in Fig.~\ref{RneckVsC02_-1}(a) and (b)), corresponding to a membrane morphology classified as tubular neck, as depicted by the dotted profile in Fig.~\ref{RneckVsC02_-1}(d). These results suggest that the extended-Helfrich model $\kappa_{12} = 0$ is able to produce vesiculation if the coating area is large enough.

\begin{figure}[htp]
\centering
\includegraphics[width=\linewidth,keepaspectratio]{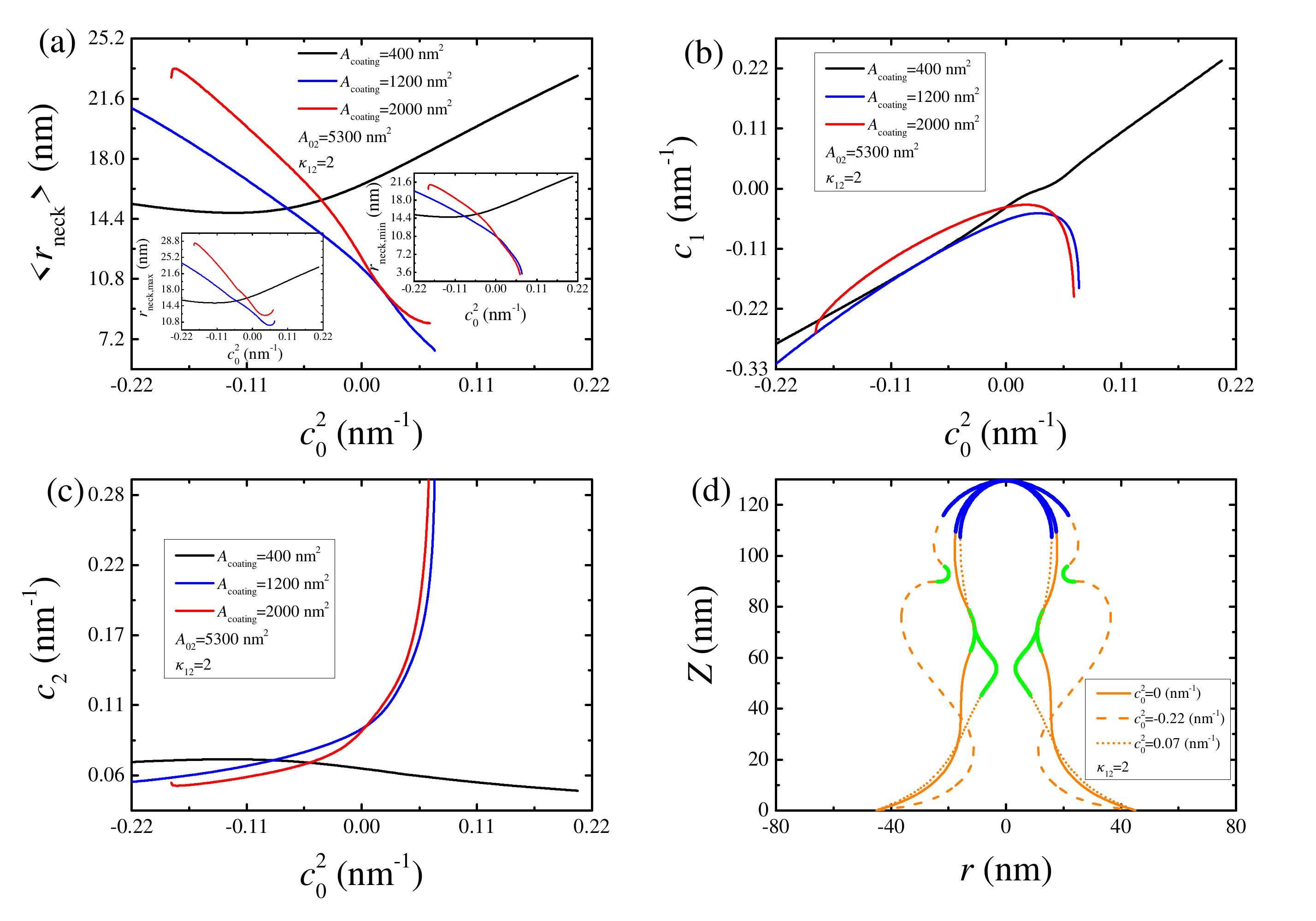}
\caption{The mean value of the neck radius $\langle r_{\rm neck} \rangle $ in (a) and the two principal curvatures $\langle c_1 \rangle $ in (b) and $\langle c_2 \rangle $ in (c), as a function of the spontaneous curvature $c_0^2$ for $\kappa_{12} = 2\kappa$. The average is taken over the protein-coated area at the side of the membrane. The three curves in (a-c) correspond to different coating areas. (d) Profile views of membrane morphologies for positive (dotted line), negative (dashed line), and zero (solid line) spontaneous curvature $c_0^2$.
}
\label{RneckVsC02_1}
\end{figure}

Subsequently, we consider the model $\kappa_{12}=2\kappa$ . As the coating area exceeds a certain value, the average neck radius $\langle r_{\rm neck} \rangle$ decreases monotonically with the spontaneous curvature $c_0^2$. In particular, the minimum neck radius $r_{\mathrm{neck,min}}$ drops to a few nanometers, indicating the occurrence of vesiculation (see the blue and the red curves in the inset of Fig.~\ref{RneckVsC02_1}(a)). The morphology of the membrane in this situation corresponds to an hourglass-shaped neck (see the dotted profile in Fig.~\ref{RneckVsC02_1}(d)), which is reflected in the large magnitude of principal curvatures $c_1$ and $c_2$ with opposite signs (the red and the blue curves in Fig.~\ref{RneckVsC02_1}(b) and (c)).

\begin{figure}[h!]
\includegraphics[width=1\linewidth]{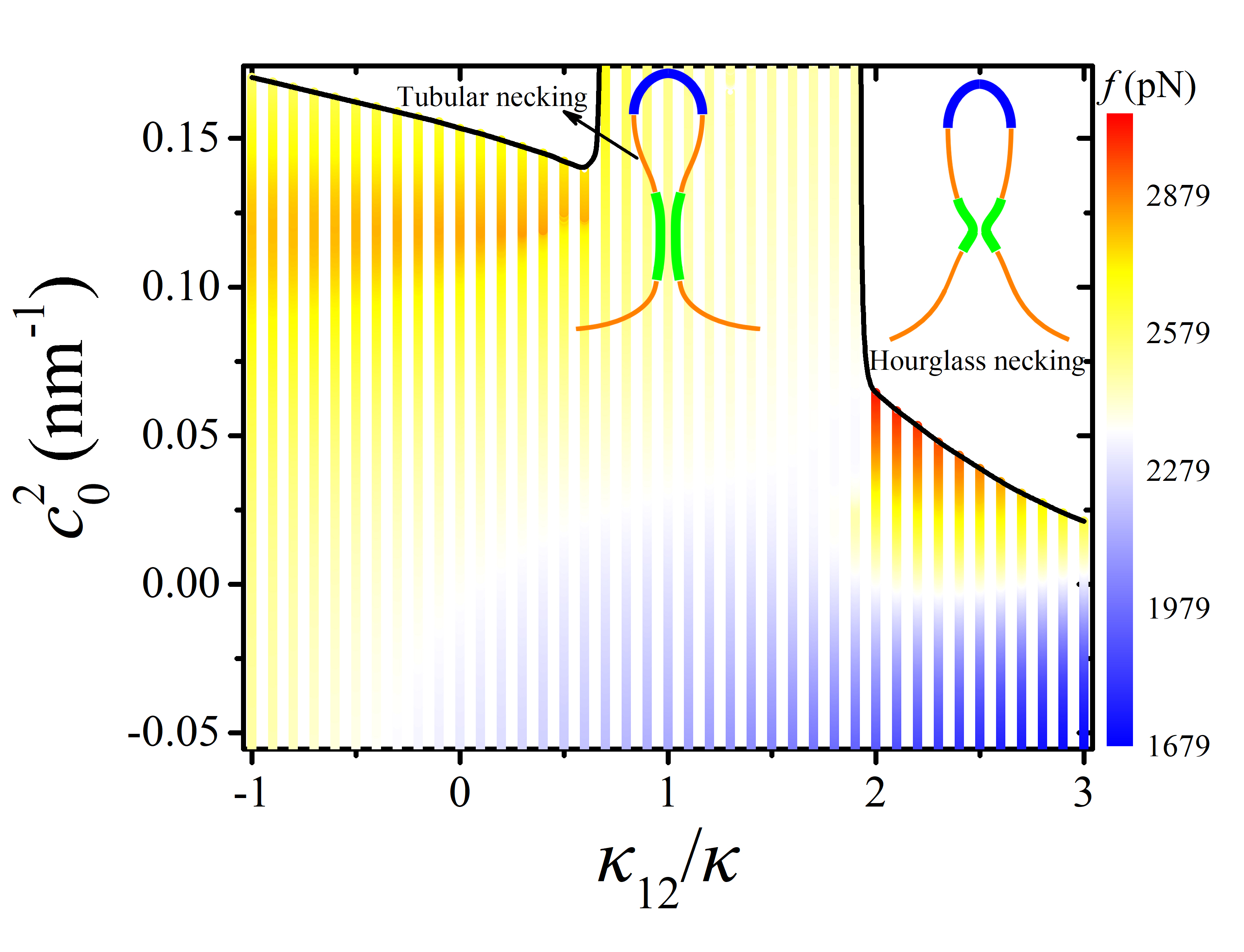}
\caption{A two-dimensional phase diagram on the ($c_0^2$--$\kappa_{12}$) plane characterizes the interrelated effects of the spontaneous curvature $c_0^2$ and the coupling constant $\kappa_{12}$ on the formation of vesicles. The colored region represents the membrane shapes that have not undergone vesiculation with a color code demonstrating the force magnitude to maintain the membrane at a tubular shape. The white regions represents the membrane shapes that have the necking radius smaller than a critical value of $5~\mathrm{nm}$, by which a vesiculation is regarded to occur. The top left corner denotes a tube-shaped necking and the top right corner denotes an hourglass-shaped necking. The coating area $A_{\rm coat}$ is $1200~{\rm nm}^2$.
}
\label{PhaseDiagram}
\end{figure}

In order to systematically investigate how membrane morphology depends on the coupling constant $\kappa_{12}$, we construct a $\kappa_{12}$-$c_0^2$ phase diagram (Fig.~\ref{PhaseDiagram}) summarizing the possible membrane morphologies. For negative and small positive values of $\kappa_{12}$, increasing the spontaneous curvature $c_0^2$ to a critical value leads to vesiculation with a tube-shaped neck (see the black curve encompassing the white region in the top left corner of Fig.~\ref{PhaseDiagram}). For large positive values of $\kappa_{12}$, vesiculation with an hourglass-shaped neck can occur if the spontaneous curvature $c_0^2$ is beyond a critical value (see the black curve encompassing the white region in the top right corner of Fig.~\ref{PhaseDiagram}). There exists an intermediate range of $\kappa_{12}$ in which vesiculation does not occur even for very large $c_0^2$.

\begin{figure}[h!]
\includegraphics[width=1.06\linewidth,keepaspectratio]{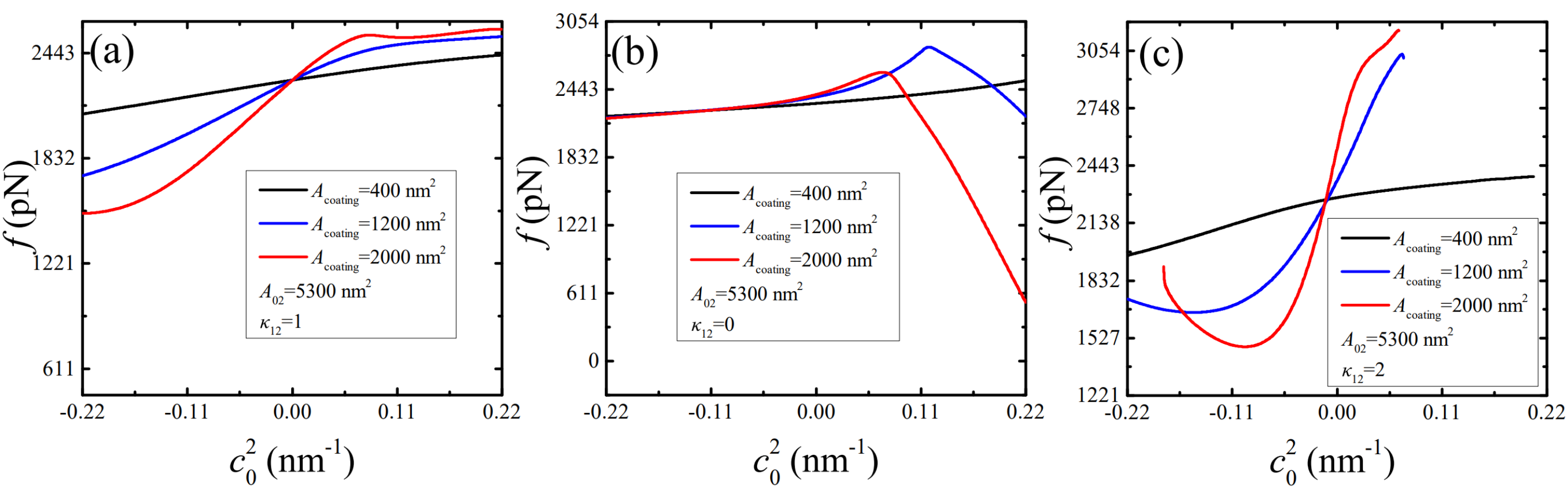}
\caption{The dependence of the pulling force $f$ on the spontaneous curvature $c_0^2$ for different coupling constants (a) $\kappa_{12}=\kappa$, (b) $\kappa_{12}=0$, and (c) $\kappa_{12}=2\kappa$. The curves could terminate at certain values of $c_0^2$ when vesiculation occurs.
}
\label{fVsC02}
\end{figure}

So far, we have shown that the coupling constant $\kappa_{12}$ neglected in most previous studies plays an important role in determining whether the anisotropic proteins at the side of the tubular membrane can drive membrane fission and generate the morphology of the membrane neck if fission occurs. Here we propose an experimental method to estimate the value of $\kappa_{12}$. It should be noted that in order to maintain the membrane at a tubular shape, a pulling force $f$ is needed to resist the membrane from being flatten since the membrane tension tends to straighten the membrane and the turgor pressure tends to push down the membrane against the cell wall. An investigation on how anisotropic proteins bound to the side of the membrane influences the force $f$ shows that the force $f$ increases with $c_0^2$ for the classical Helfrich model (see curves in Fig.~\ref{fVsC02}(a)). In contrast, for the extended-Helfrich model $\kappa_{12} = 0$, the force exhibits a gentle increase followed by a decrease with the increase of $c_0^2$ (see the blue and the red curves in Fig.~\ref{fVsC02}(b)). As for the model $\kappa_{12}=2\kappa$, a sharp increase in the force $f$ is accompanied with a small increase of the spontaneous curvature $c_0^2$ (see blue and red curves in Fig.~\ref{fVsC02}(c)). The colored region in Fig.~\ref{PhaseDiagram} demonstrates how the force depends on the combination of the coupling constant $\kappa_{12}$ and the spontaneous curvature $c_0^2$. In the intermediate range of $\kappa_{12}$ ($0.7 \kappa < \kappa_{12} < 1.9 \kappa$), the force $f$ is almost independent of $c_0^2$. For large $\kappa_{12}$ ($\kappa_{12} < 0.7 \kappa $), the force $f$ has a sharp increase with $c_0^2$. While for small and negative $\kappa_{12}$ ($\kappa_{12} < 1.9 \kappa$), the force $f$ shows a nonmonotonic dpendence on $c_0^2$.

Pulling a membrane tube from a giant liposome with optical tweezers is a common \textit{in vitro} experiment to determine the membrane tension and the membrane bending rigidity. When the tube is formed, by flowing the proteins of interests into the solution, membrane could be gradually bound with the proteins at the lateral side of the tube and the spontaneous curvature is expected to increase with the enrichment of the proteins. Measuring the force to maintain the membrane at a tubular shape in response to the growth of the protein coat, checking a figure like Fig.~\ref{PhaseDiagram}, and reading how the force depends on the protein concentration, we can have an estimation of the range of the coupling constant for that type of proteins.

BAR proteins have been proposed to be an active player in membrane fission during the late stage of endocytosis in yeast cells. In particular, the crescent-shaped N-BAR proteins have a typical radius of $10\mathrm{nm}$ and is able to induce membrane tubulation of the same radius when the density is high enough. We have shown in the phase diagram of Fig.~\ref{PhaseDiagram} that the coupling constant $\kappa_{12}$ determines whether membrane fission could happen upon increasing the spontaneous curvature $c_0^2$. For strong coupling ($\kappa_{12}>2\kappa $), a small spontaneous curvature $c_0^2$ ($>0.05~\mathrm{nm^{-1}}$) generated by the binding of BAR proteins is enough to induce vesiculation. However, for small positive and negative values of $\kappa_{12}$, a very large spontaneous curvature $c_0^2$ ($>0.13~\mathrm{nm^{-1}}$) is needed to induce vesiculation. As it is known that N-BAR proteins have a curvature of $\approx 0.1~\mathrm{nm^{-1}}$ (smaller than $0.13~\mathrm{nm^{-1}}$ but greater than $0.05~\mathrm{nm^{-1}}$), our results therefore suggest that N-BAR proteins are able to actively induce membrane fission not via tubular necking but via hourglass necking.

As a result of the high turgor pressure inside yeast cells, maintaining the membrane at a tubular shape needs to generate a very large force. Actin polymerization is assumed to provide the force. However, based on the copy number analysis of actin filaments, polymerization alone seems unable to generate enough force~\cite{M.Rui2021,J.Berro2010}. We have found that anisotropic proteins with a coupling constant $\kappa_{12} = 0$ could significantly reduce the force to maintain the membrane at a tubular shape from $2000~\mathrm{pN}$ to $600~\mathrm{pN}$. This result provides a new perspective to explain the large difference between the required force and the actual force generated by actin polymerization.


In summary, we study the physics behind vesiculation phenomena via anisotropic proteins bound to the side of a tubular membrane during endocytosis.
It is found that the classical Helfrich model is incapable of explaining vesiculation.
Anisotropic spontaneous curvatures based on the extended-Helfrich model are needed to drive membrane fission. Depending on the type of anisotropic curvatures,  the membrane tube could undergo an tubular necking or an hourglass necking. Furthermore, we suggest an experimental method to distinguish the type of anisotropic curvatures of a protein by comparing the force to maintain the membrane at a tubular shape in the presence and absence of the proteins.

\section{ACKNOWLEDGMENTS}
We acknowledge financial support from National Natural Science Foundation of China under Grants No. 12147142, No. 11974292, No. 12174323, and No. 12004317, Fundamental Research Funds for Central Universities of China under Grant No. 20720200072 (RM), and 111 project No. B16029.



\end{document}